\documentclass[aps,prb,superscriptaddress,twocolumn]{revtex4-2}
\usepackage[english]{babel}
\usepackage{graphicx}

\usepackage[letterpaper,top=2cm,bottom=2cm,left=3cm,right=3cm,marginparwidth=1.75cm]{geometry}

\usepackage{amsmath}
\usepackage{graphicx}
\usepackage[colorlinks=true, allcolors=blue]{hyperref}

\newcommand{\tsc}{\ensuremath{C_\mathrm{TS}}}
\newcommand{\stc}{\ensuremath{C_\mathrm{Str}}}

\newcommand{\ZTot}{\ensuremath{Z_\mathrm{Tot}}}

\newcommand{\dtip}{\ensuremath{d_\mathrm{tip}}}
\newcommand{\delf}{\ensuremath{\Delta_\mathrm{eff}}}

\begin{document}

\title{Adding Radio Frequency Capabilities to a millikelvin Scanning Tunneling Microscope}

\author{Jonathan Marbey}
\affiliation{Department of Physics, University of Maryland, College Park, MD, USA}

\author{Michael Dreyer}
\affiliation{Department of Physics, University of Maryland, College Park, MD, USA}

\author{R.E. Butera}
\affiliation{Laboratory for Physical Sciences, College Park, MD, USA}

\begin{abstract}
We present a simple home made solution enabling in-situ RF reflectometry measurements with a millikelvin scanning tunneling microscope (mk-STM). The additions described below were made using RF best practices following similar detection schemes commonly employed in the quantum information science (QIS) community. Using a Niobium STM tip to form a superconductor-insulator-normal metal (SIN) tunnel junction, the evolution of coherence peaks at the SC-gap edge are carefully measured to characterize the RF losses and electron temperature. We further identify impedance matching as a crucial factor to achieve high sensitivity in the reflectometry by tuning the tip-sample capacitance as a function of approach distance. As a demonstration of this capability, we measure a 50x50 nm$^2$ area of island features that have been condensed onto the surface of a gold single crystal. Position dependent reflectometry losses allow us to image island sizes down to a total surface area of 5 nm$^2$ given our current sensitivity. 
\end{abstract}

\maketitle
 
\section*{Introduction}
Scanning tunneling microscopy (STM) typically operates within a bandwidth of just a few kHz. Expanding the bandwidth into the radio frequency (RF) range enables a variety of applications -- such as noise measurements, electron spin resonance (ESR), scanning on insulating materials, or measuring tip-sample capacitance -- which have been successfully demonstrated across different systems and environments\cite{RFSTMstefan, nat2007, ESRlutz, steveSTM}. The specific frequency regime and instrumentation employed within each of these techniques is largely determined by the specific application. 

Our particular application of RF-STM focuses on sensing local capacitive or resistive variations within either conductive or semi-conductive materials. This is achieved through direct measurement of the local impedance using a resonant tank circuit (at $\sim$300 MHz) that is integrated into our home-built millikelvin (mK) system \cite{michael}. The tank circuit, described in more detail below, is composed of a simple hand-wound inductor that is connected in-series to the STM tip. In this configuration, reflectometry measurements of the tank circuit response allows us to probe local permittivity variations at sub-nanometer scales. 

Although scanning capacitance techniques are well-established within atomic force microscopy (AFM), their implementation in STM remains relatively scarce \cite{steveSTM, nat2007}. The most common motivation to pursue RF-STM is to increase the bandwidth of the feedback control to allow for faster scanning and approaching. However, the application presented herein is more closely aligned with microwave impedance microscopy (MIM) \cite{MIM_UCSD, MIM_REV} in low temperature environments, but with the added benefit of enhanced spatial resolution afforded by STM. Most importantly, we draw particular inspiration from the quantum information science (QIS) community where low-power dispersive sensing schemes are employed in quantum dot devices to improve charge-state read out \cite{reviewReflect}. In this context, the signal to noise ratio (SNR) is augmented by measuring sub-aF changes to the so-called "quantum capacitance" of a quantum dot, rather than low frequency (near-DC) conductance associated with single electron charge sensors \cite{Tahan2019}. 

In practice, integrating dispersive sensing techniques into existing STM systems is challenging as parasitic sources of capacitance and inductance are difficult to isolate in scanning geometries. In this paper, we present a straightforward approach to incorporating a tank circuit design into a simple STM tip-plate, while discussing limits to sensitivity. Despite our current constraints, we demonstrate the capability to detect extremely weak variations in the reflectometry (on the order of mdB) of surface features as small as 5 nm$^2$ with sufficient spatial resolution.

The paper is organized as follows. We first outline the basic principle of operation associated with reflectometry measurements at RF frequencies in Section I. Section II details the RF components, instrumentation and wiring that enables RF-STM. Sections III and IV address technical parameters associated with RF performance, including transmission to the STM tunnel junction, electron temperature, and tip-position dependent impedance matching. As a practical demonstration of this technique, Section V summarizes near-field reflectometry measurements of sub-monolayer island-like features of ice on Au(111). 

\section*{I. Principle of operation}

\begin{figure*}[htp]
\centering
	\includegraphics[width=1\textwidth]{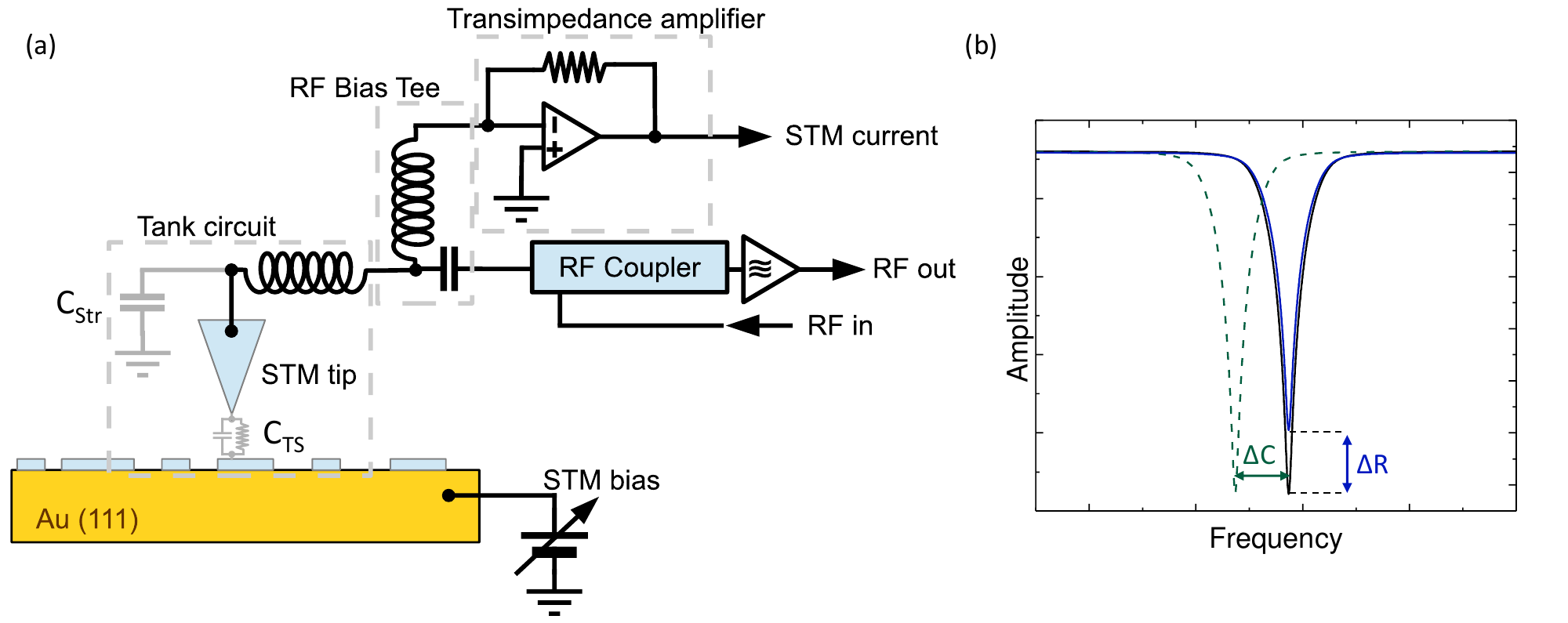}
	\caption{(a) Schematic of the tank circuit integration into the mK-STM with relevant RF components as described in the main text. (b) Illustration of how capacitive variations (green) and resistive (blue) variations affect the amplitude associated with the reflectometry of the tank circuit.}
 \label{Fig:overview}
\end{figure*}

 We begin with the description of a simple series connected LCR resonator with the specific goal of quantifying how variations in circuit parameters affect the reflectometry of a STM junction. From a lumped element perspective, STM geometry can effectively be modeled as an irregularly shaped capacitor with an adjustable-sized vacuum gap separating the tip and sample. For a spherical tip approximation, the change in 'tip-sample' capacitance, \tsc, can be roughly described by $ d\tsc/dz \sim 1/z$ \cite{Hou}. However, as the tip approaches tunneling distance (z $<$ 1 nm), the details of the STM tip shape, along with local surface structures and/or charge distributions will cause the capacitance to deviate from this simple relation. 
    
    In an ideal scanning reflectometry experiment, the height is fixed at a short tip-sample distance (i.e. 'z' is held constant and  $ d\tsc/dz =0$), such that \tsc\ is only sensitive to local surface variations (e.g. in presence of lateral charge fluctuations). By combining the STM tip with a tank circuit, the total complex impedance can be described in the familiar series summation of individual components: $\ZTot= Z_C + Z_L + R$, with $Z_L$ being given by an in series inductance, forming an $LC$ resonator with series resistance $R$.  Here, in addition to \tsc, $Z_C$ must also consider contributions from any source of stray parasitic capacitance, \stc, such that the total capacitance $C=\tsc+\stc$. In practice, typical values of $\stc$ are on the order of 0.5-1 pF whereas $\tsc$ can range from aF to fF, depending on conditions related to both tunneling capacitance and the sample environment \cite{Tahan2019, Hou}. For the purposes of our current discussion, we therefore choose to treat the stray capacitance as an umbrella term that encapsulates any other source of capacitance other than $\tsc$, which may need to be added in parallel or series for any particular component. 
    
    The real contributions to the impedance, given by $R$, are dominated by standard RF components with an ideally matched 50 $\Omega$ impedance, but also contains parallel sources of free space impedance ($\sim$377 $\Omega$) and tunnel resistance (typically many $M\Omega$ to a few $G\Omega$ which varies exponentially pending the tip-sample distance \cite{nat2007}). Analogous to the description of capacitance, when z is fixed such that $dR/dz = 0$,  variations in $R$ are also dominated by local structure.

As we are primarily interested in small changes to $\ZTot$ over nanometer-length scales across a sample of interest, it is necessary to discuss how variations in circuit parameters affect the sensitivity of the tank circuit. From the assumptions described above, the reflected phase and amplitude are conventionally derived from the magnitude and complex argument of the well-known reflection coefficient: 

\begin{equation}
	\Gamma(\omega)=\frac{ Z_\mathrm{Tot}(\omega) - Z_0}{ Z_\mathrm{Tot}(\omega) + Z_0} \label{eq:gamma}
\end{equation}

Here, $Z_0$ is determined by the co-axial line impedance which is conventionally constrained to 50 $\Omega$ in most RF applications. For small capacitive changes to the STM tank circuit, we define sensitivity with respect to variations in the total $C$ \cite{morton2018}:

\begin{equation}
	\big | \frac{\partial \Gamma}{\partial C} \big |\Delta C =\frac{2Z_0 R}{(R+Z_0)^2} Q \frac{\Delta C}{C}
	\label{eq:dgdc}
\end{equation} 

Likewise, resistive variations can be defined by:
\begin{equation}
	\big | \frac{\partial \Gamma}{\partial R} \big |\Delta R =\frac{2Z_0}{(R+Z_0)^2}\Delta R
	\label{eq:dgdr}
\end{equation}

Where we have substituted the inductance $L$ with the conventional LCR quality factor  $Q =\frac{1}{R}\sqrt{\frac{L}{C}}$, and represent the in-series resistance 'R' at resonance where $f = f_0$, (i.e. $f_0=\frac{1}{2\pi\sqrt{L\cdot(\stc+\tsc)}}$). 

By inspection of Eq.~(\ref{eq:dgdc}), it is easy to see that sensitivity to $\Delta C$ is optimized for large values of $Q$ and minimal parasitic  contributions \stc to the total  capacitance $C$ . Meanwhile, sensitivity to variations in resistance are maximized when the total resistance $R$ is minimized in Eq.~(\ref{eq:dgdr}). Interestingly, the ideal choice of resistance for sensitivity in  Eq.~(\ref{eq:dgdr}) does not hold for Eq.~(\ref{eq:dgdc}), where the extra factor of R instead requires $Z_0 = R$. 
These familiar resonator requirements extrapolate well to the situation concerning our integrated tank circuit described below when neglecting stray sources of impedance. However, it is important to emphasize that while we are only interested in measuring $\tsc$ and R, the reflectometry is indeed sensitive to any variation in capactive or inductive contributions. As will be addressed below, this sensitivity causes significant complications in the context of STM, as local variations in $\stc$ (e.g. as a function of vertical tip position) have the potential to swamp $\tsc$.

\section*{II. Instrumentation and Measurement Scheme}

\begin{figure*}[ht]
	\centering
	\includegraphics[width=1\textwidth]{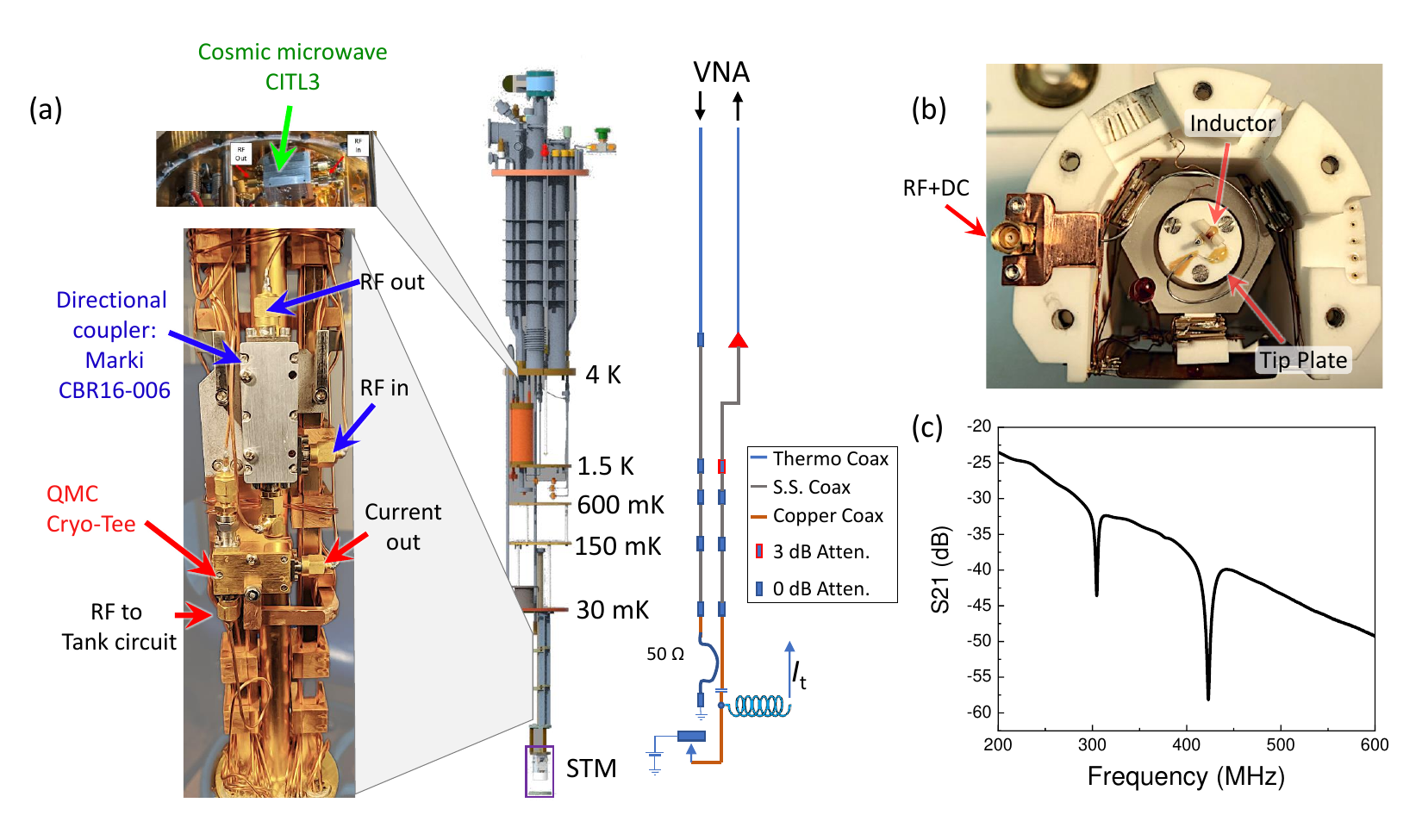}
	\caption{(a) Outline of RF scheme integrated into the DR. The majority of the components, described in the main text are mounted below the mixing chamber (30 mK) stage. The associated input/output ports are labeled via the blue and red arrows. The coaxial RF lines are shown parallel to the DR unit. Blue lines from room temperature to 4K stage represent thermocoax, the 4K to MXC lines are silver plated stainless steel coax, MXC to STM cold finger are copper coax. (b) Modified STM tip plate that houses the tank circuit situated within the STM walker assembly. Here, the inductor is hand wound around an aluminum oxide post, which is kept fixed to the plate via epoxy. Electrical contact to the tip holder and coax line is made via spot weld, while the contact to the center conductor of the coax is made via silver paste. The coax is ultimately fed into an SSMC port. (c) Room temperature reflectometry of the tank circuit, measured as S21 with the cryogenic amplifier only.} 
 \label{Fig:layout}
\end{figure*}

In practice, $\Gamma(\omega)$ is not measured directly, but can instead be reconstructed from the amplitude $S_\mathrm{11} = \log_\mathrm{10}|\Gamma(\omega)|$ and phase $\Phi = tan^{-1 }(\frac{Im(\Gamma(\omega)}{Re(\Gamma(\omega)})$ using a network analyzer. The associated RF measurement scheme is summarized in the diagram shown in Fig.\ \ref{Fig:overview}(a). Here,  simultaneous measurement of the STM current and reflectometry is possible due to the combination of a cryogenic bias tee and directional coupler. The tank circuit itself is formed from the combination of a simple inductor connected in series to the STM tip, and a capacitive coupling between the tip and sample of interest. The two most obvious sources of capacitance, $\tsc$ and $\stc$ (as described in Section I), are labeled in the vicinity of the tip. The resultant geometry can thus be approximately modeled by an effective LCR model with total $C=\tsc+\stc$. In the most common mode of operation, the tank circuit is fixed to its resonant frequency while the STM bias or tip position is varied. As a result, the net output from the tank circuit contains both a DC-STM tunneling current component and an AC reflectometry component. 

The RF scheme described here was installed onto a pre-existing mK-STM that is housed within a wet oxford kelvinox 400HA dilution referigerator (DR) \cite{markthesis}. The original filtering scheme for all RF connections constrained the operational bandwidth to $\sim$20 kHz such that incoming thermal radiation at higher frequencies could be sufficiently attenuated. As a result, a number of previously established design decisions impose strong limitations on the achievable bandwidth of our set-up. With this in mind, all RF modifications were done such that no major overhaul to the instrumentation was required to achieve adequate sensitivity.  Where necessary, we acknowledge these limitations when describing the instrumentation, and leave a number of items as a subject for discussion in the future work section at the end of this paper. 

The S-parameter measurement is performed using a 2 port network analyzer (Keysight PNA N5222B). Adjacent stages of the DR are coupled via 0 dB attenuators (QMC Cryo-attf), except for a single 3 dB attenuator on the return path of the 1.5 K stage which provides additional thermal anchoring. Each cryogenic attenuator is mounted to its respective stages via copper brackets so that the outer housing can be effectively thermalized. A combination of thermo-coax, low-loss silver-plated stainless steel coax (CryoCoax BCB028), and generic 0.047" copper coax are employed to transmit/receive RF signals to the tank circuit (see Fig.\ \ref{Fig:layout}a for details). Stainless steel coaxial cabling is employed to reduce thermal coupling between adjacent stages down to the mixing chamber (MXC), while copper coax is used \textit{below} the MXC to maximize coupling. It is also important to note that in this current configuration, the thermo-coax spanning the room temperature feed-thrus to the 4K stage accounts for the largest source of RF loss. In the context of mK-STM, thermo-coax is commonly employed as a distributed RC filter due to its higher resistance and capacitance\cite{markthesis}. As such, the pre-existing thermocoax in our DR is employed in these measurements out of convenience, and thus imposes the hardest limit on our achievable bandwidth. Future iterations of our RF scheme will replace this path entirely. 

To measure the resonator reflectometry, we have chosen to employ entirely commercial components, each of which were purchased without modification and incorporated into the mK-STM system described in ref.\cite{michael}. A more detailed description follows below according to an outline of the implementation in Fig.\ \ref{Fig:layout}(a). The combined DC/RF signal is fed from the output of the tank circuit via 30 cm of 0.047" Cu coax up to a cryogenic bias tee (QMC-CryoTee-0010), which is thermally anchored below the mixing chamber stage via a cold finger extension. This bias tee effectively parses the tunneling current and RF signals such that they are directed to their respective measurement paths. Notably, the bias tee has the effect of giving rise to a measurable cross talk between the DC and AC channels. This cross-talk manifests as a small increase in noise in the STM tunneling current when power is applied to the RF input. Likewise, the signal from the tunneling current also leaks into the AC branch of our detection circuit. We compensate for these cross-talk effects using a simple decorrelation procedure described in the Appendix. The reflected RF signal is further resolved from the input signal via a directional coupler (Marki CBR16-006) before proceeding thru a +33 dBm cryogenic HEMT-based amplifier (Cosmic Microwave Technology CITLF3). We note that the Marki directional coupler is not specifically intended for use at cryogenic temperatures as various internal components undergo a superconducting phase transition below a nominal value of $\sim$5\ K. At these lower temperatures, the loss through the coupler arm is actually reduced as described at the end of section III.  However, by reducing the thermal coupling to the MXC and applying moderate amounts of RF power, we are able to recover the specified behavior. The unideal choice of this particular directional coupler was made based on the tight space constraints towards the lower stages of the DR. To our knowledge, there are no commercially available cryogenic directional couplers or circulators that operate below 1 GHz. 

The tank circuit, which is mounted onto a modified STM tip plate, is comprised of a hand-wound inductor consisting of 12 turns of copper-clad NbTi wire (Supercon Inc. Wire Type: T48B-M) wrapped around an aluminum oxide core of diameter 1.6 mm dimensions (see Fig.\ \ref{Fig:layout}(b)). These dimensions correspond to an ideally calculated inductance of $L$ = 470 nH assuming a simple solenoid. Super-conducting wire is used here to reduce the in-series resistance at the base temperature of the DR. In combination with the total tip-sample stray capacitance, the tank circuit yields a resonant frequency of around $\sim$300 MHz at room temperature, which corresponds to $C = \stc \approx$  0.60 pF when the tip is relatively far from the sample (i.e $>$ 30$\mu$m, $\tsc$ = 0), Fig.\ \ref{Fig:layout}(c). We note that exact resonance frequency is highly dependent on tip-sample geometry, and varies from cool down to cool down within a few MHz frequency window. These details are addressed in Section IV. A frequency sweep of the network analyzer also reveals a nearby second resonance centered at 423 MHz which can likely be attributed to a stray capacitance parallel to the inductor. At the base temperature of the DR, the quality factor of this second resonance reduces dramatically and yields very little sensitivity to variations in sample topography (e.g. performing measurements as described later Section V). The resultant baseline in Fig.\ \ref{Fig:layout}(c) shows a total loss of ~30 dB in the vicinity of the resonance of interest at room temperature. In all subsequent measurements shown below, we compensate for this loss with a room temperature RF amplifier from mini circuits (ZFL-1000LNB+).   
\section*{III. Power calibration from S-I-N Tunneling}

\begin{figure*}[ht]
	\centering
	\includegraphics[width=1\textwidth]{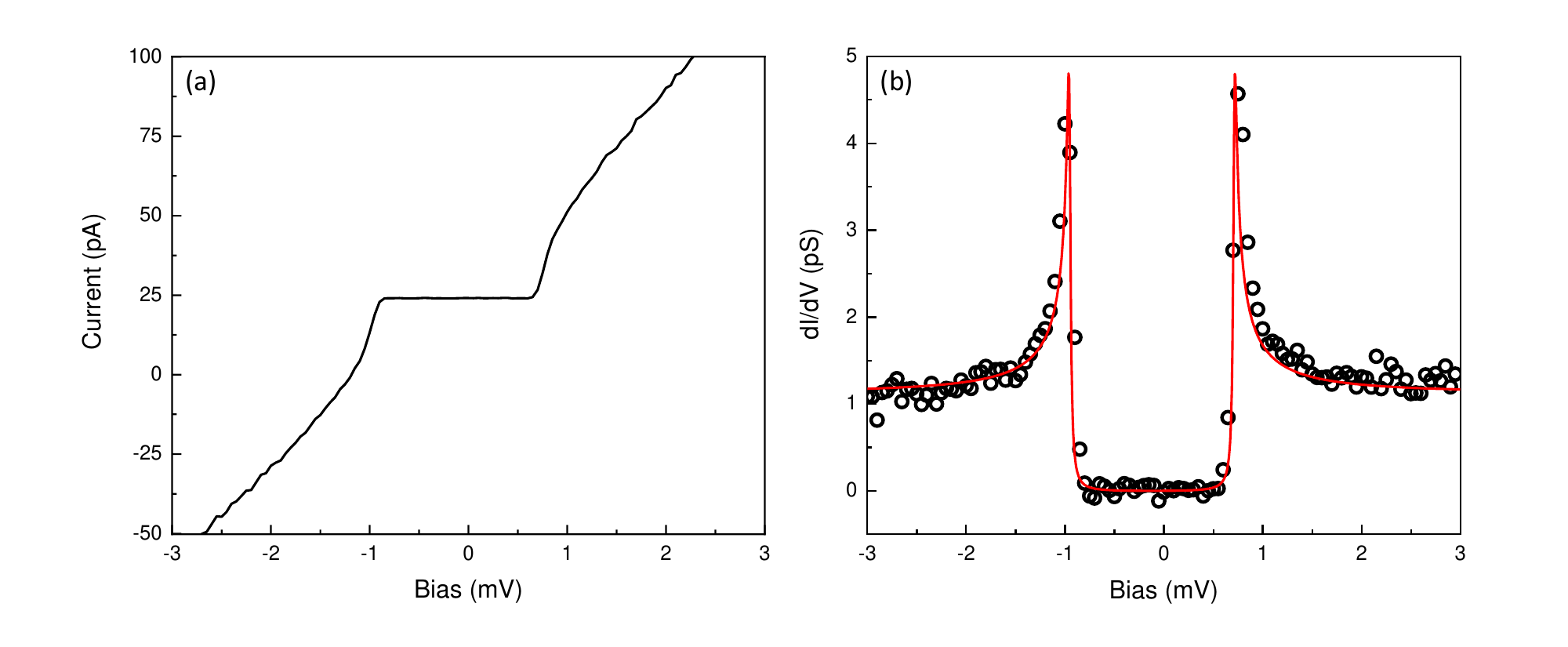}
	\caption{(a) I(V) and (b) dI/dV with zero applied RF power to the tank circuit. The data are obtained by turning off the STM feedback control, fixing the tip to a constant height, and varying the tip-sample bias. Both sets of data consist of 20 averaged spectroscopy curves. The red curve is obtained from a fit to Eq. 3.}
 \label{Fig:SIN}
\end{figure*}
 
In conventional DR set-ups utilizing different combinations of coax, attenuators and other RF components, it is standard practice to integrate a cryogenic switch in the vicinity of the sample stage such that: i. power can be directly measured at the sample of interest and ii. proper de-embedding techniques can be employed \cite{de-embed}. Unfortunately, our set up has a rather small foot print below the mixing chamber stage and is therefore unable to accommodate the size of any commercially available switches. An alternative method to measuring RF losses near the tip can instead be achieved using a Nb STM tip on top of a metal sample, where the coherence peaks at the vicinity of the gap edge can be used as a marker to measure electron temperature\cite{MichaelTipTemp}. Additionally, application of RF to the tunnel junction via the bias tee super-imposes a periodic voltage modulation that effectively splits the coherence peaks at voltage scales proportional to the applied RF power at the junction. As such, the superconductor-insulator-normal metal (SIN) junction afforded by our measurement geometry provides an effective method to characterize both the total RF losses associated with the tank circuit integration and the effective tip temperature in the presence of RF power. 

The Nb tip employed in the measurements is chemically etched from a 0.25 mm outer diameter wire (purity 99.99\%) in a solution of 4M KOH \cite{NbTipEtch}. This preparation yields tip curvatures typically on the order of 100-1000 nm. After etching a fresh Nb tip, it is vital to load and pump down to base pressure as quickly as possible to minimize oxide and nitride formation at the tip surface. The Au(111) sample is cleaned in an ultra high vacuum (UHV) preparation chamber by simultaneously sputtering and heating to 400\textdegree C. After preparation, the sample is inserted into the STM via a series of magnetic transfer arms, which is pre-cooled to the DR base temperature of $\sim$30 mK according to calibrated thermometry \cite{michael}. 

Preliminary measurements of tunneling current as a function of tip-sample bias, $I(V)$, and conductance, $G(V)$= $dI/dV$, in the absence of applied RF power, shown in Fig.\ \ref{Fig:SIN} reveal spectra characteristic of S-I-N tunneling. The resulting conductance data is well described by the phenomenological Dynes equation \cite{dynes}:

\begin{equation}
	p(E, i\gamma) = \frac{E-i\gamma}{[(E-i\gamma)^2-\Delta^2]}
	\label{eq:ampsens}
\end{equation} 
where the BCS density of states $p(E, i\gamma)$ is expressed in terms of the gap width $\Delta$ and an imaginary contribution $\gamma/\hbar$ being the quasiparticle relaxation rate. As such, $\gamma$ effectively parameterizes the thermal broadening in the vicinity of the coherence peaks at the gap edges. A fit to Eq.~\ref{eq:ampsens}, given by the red curve in Fig.\ \ref{Fig:SIN}(b), yields an effective tip temperature of 171 mK, corresponding to $\gamma$ = 0.0148 (124) meV with $\Delta$ = 0.831(3) meV. Though the fit error here is quite large, we note that this effective temperature is in line with previous measurements of $T_{eff}$ = 184 mK \cite{michael}. From this, it is safe to conclude that the addition of RF components to the original set up does not have a drastic effect on total cooling power of the DR. 

\begin{figure*}[ht]
	\centering
	\includegraphics[width=1\textwidth]{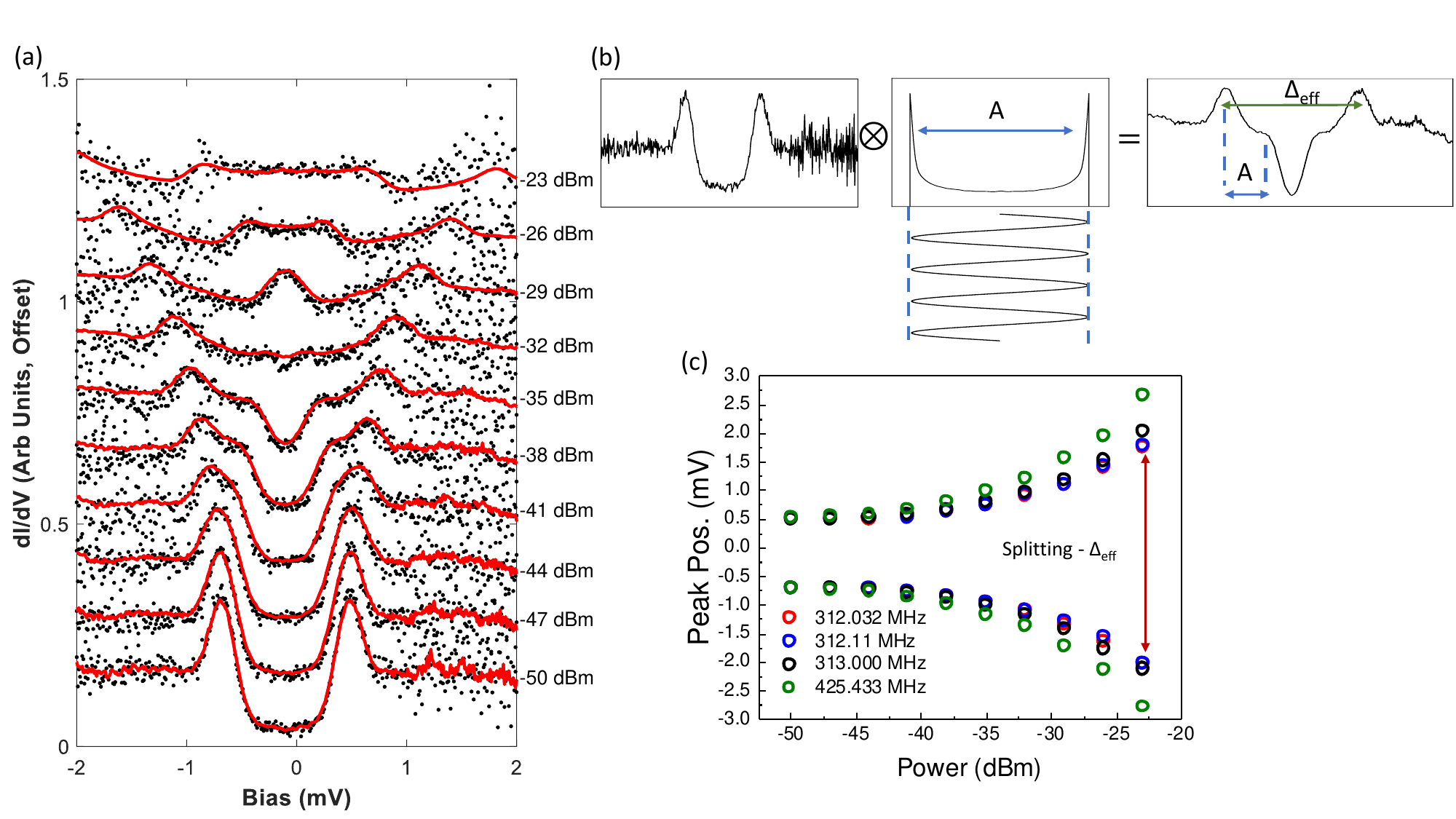}
	\caption{(a) Stacked conductance data at $f_0$ = 312.032 MHz (black points) with each successive curve corresponding to an increase in the RF port power by increments of 3 dB. The overlaid red curves are formed from the convolution described in the main text, which is graphically summarized in (b). (c) Splitting of the modulated coherence curves as a function of power for four  different frequencies: on resonance (312.032 MHz), 3dB down (312.111 MHz), far off resonance (313 MHz) and the secondary resonance (425.433 MHz). The splitting is measured via peak to peak distance, as labeled in the rightmost curve in (b).}
 \label{Fig:pdep}
\end{figure*}

Fig.\ \ref{Fig:pdep}(a) shows the power dependence of the dI/dV spectra with the tank circuit tuned to resonance. After application of relatively low RF port power of -50 dBm (bottom most curve), the conductance peaks appear to immediately broaden, corresponding to a tip temperature of 490(99) mK with an associated superconducting gap $\Delta$ = 0.481(7) meV. However, despite this obvious large difference in $\Delta$ and $\gamma$, the thermometry associated with the mixing chamber and STM stages of the dilution refrigerator showed little variation. The lack of a measurable temperature deviations suggests that a significant contribution to the apparent broadening in the -50 dBm conductance data is due to tip contamination. Such contamination can either arise from: i. coulombic attraction of mobile or weakly bound defects on the sample surface to the STM tip, or ii. unavoidable atomic changes to tip volume after scanning. Though the exact condition of the tip makes it quite difficult to disentangle the exact electron temperature difference from the case of 'power off' vs 'power on' data, the trace recorded at -50 dBm nevertheless serves as a suitable baseline to compare against dI/dV traces recorded at higher port powers in Fig.\ \ref{Fig:pdep}(a).
	 
As the RF port power is further increased in steps of 3dBm (equivalent to iteratively doubling the port power), the coherence peaks symmetrically split. This splitting is most obvious starting at -38 dBm, where a broad shoulder centered roughly at 0.25 mV emerges. From here, as the port power is raised further, the splitting between the shoulder and the apparent peak gradually increases until the shoulders on either side of the gap cross over each other near 0 mV. The top 3-most traces reveal the emergence of yet another peak in the conductance, centered with respect to the superconducting gap, which begins to broaden up to -23 dBm. Above this power level, the spectra become too broad to distinguish from the inherent noise of the measurement.

Though this behavior of the conductance data might appear complicated at first glance, it can be simply explained by the RF modulation that is superimposed onto the DC tip-bias voltage via the cyrogenic bias tee  (\cite {MichaelTipTemp, Heinrich,  TipTempNotNb}): 

\begin{equation}
	M(V) = \int_{0}^{T} G_\mathrm{-50dBm}(V+\frac{A}{2}\sin(\omega \tau) ) \,d\tau \
	\label{eq:ampconv}
\end{equation} 

\begin{figure*}[ht]
	\centering
	\includegraphics[width=1\textwidth]{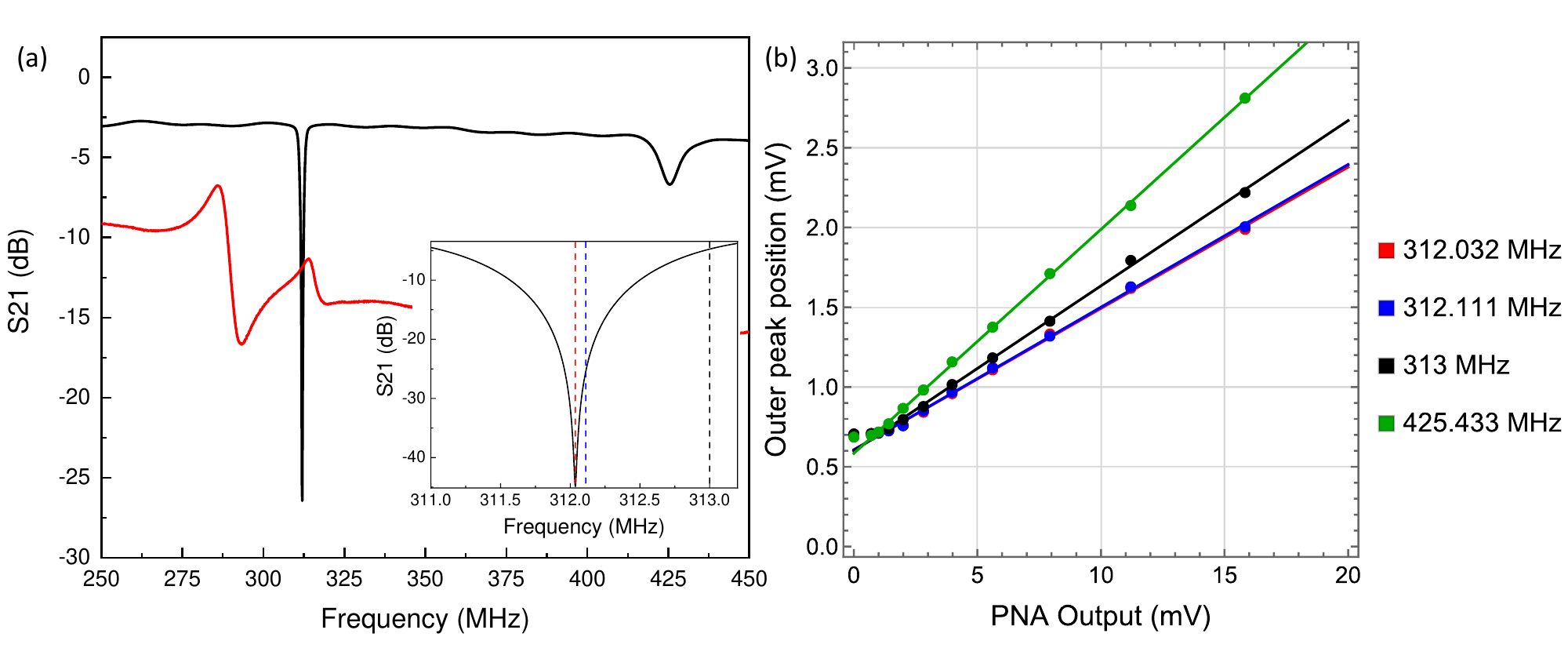}
	\caption{(a) Frequency traces of the tank circuit at base temperature with port powers 16 dBm (black) and -30 dBm (red). These applied powers correspond respectively to the normal and superconducting temperature regimes of the directional coupler. The inset shows a finer sampled frequency trace in the vicinity of the main resonance in the tank circuit. The colored dashed lines correspond to the frequencies sampled in the power dependence (b) Voltage positions of the modulated coherence peaks (see Fig. \ref{Fig:pdep}c) as a function of the output of the network analyzer. Note that the x axis is in mV, ranging from 0.71 mV (-50 dBm) to 15.83 mV (-23 dBm). The slopes of the red, blue, black and green linear traces are respectively 0.0886(12), 0.0894(12), 0.1038(12), 0.1404(8).}
 \label{Fig:pdepLin}
\end{figure*}

The modulated conductance at elevated RF power, M(V), is formed from a time averaged convolution taken over a single period $T$ of the conductance at a low base RF power, $G_\mathrm{-50dBm}(V)$, with a modulating sine wave of amplitude A. The calculation of M(V) is graphically represented in Fig.\ \ref{Fig:pdep}(b), which illustrates the convolution of a modulated spectroscopy curve at -33 dBm port power. Here, the time averaged sine wave is calculated from the distribution over a single period, with the peak to peak distance given by the amplitude A. 
Fig.\ \ref{Fig:pdep}(c) plots the conductance peak splitting \delf\ with respect to the PNA RF port power converted to mV RMS at three frequencies in the vicinity of the primary resonance (here, $f_0$ = 312.032 MHz) in addition to the secondary resonance at 425.433 MHz. As such, the rate at which \delf\ increases with respect to applied port power provides an effective method to measure the RF amplitude applied to the S-I-N tunnel junction. This rate is obtained from fits to the linear part of the  trend shown in Fig. \ref{Fig:pdepLin} (b). From the fitted slope, we can define  RF power loss as \mbox{(1 - slope)}. For example, a slope of 0.088 for $f = f_0$ corresponds to a power loss of 91.2$\%$. To confirm the accuracy of this fit, each trace in Fig. \ref{Fig:pdep}(a) is recreated according to a rescaled amplitude 'A' in Eq. \ref{eq:ampconv}\ based on the fitted splitting in Fig.\  \ref{Fig:pdepLin}(b). The calculated tunneling spectra (in red) replicate the modulated spectroscopy remarkably well, even for the case of lower port power which deviates from the linear trend. This calculation therefore presents a very sensitive method to infer the rectified amplitude at the tunnel junction by replicating the appropriate peak-peak distance in the spectroscopy.

Naively, one would expect that the amplitude modulation at the tunnel junction becomes stronger in the vicinity of the tank circuit resonance (i.e. where the resonator absorbs the most power, $f = f_0$). However, we observe the opposite effect; the fitted coherence peak positions shown in Fig.\ \ref{Fig:pdepLin}(b) reveals that the power delivered to the junction becomes more attenuated closer to $f_0$. This behavior follows well out to the 2nd resonance centered at 425 MHz where the total power loss jumps from 91$\% $ (at 312.032 MHz) to 86$\%$ (see Figure caption). We attribute this frequency dependent loss trend to additional parasitics within the tank circuit. As such, the trend suggests that a high-pass like cut off dominates the frequency dependence we observe in the amplitude modulation at the junction. Here, the most obvious source of capacitance would involve a coupling between the STM tip and sample, though as discussed in the proceeding section, the sources of various parasitics are hard to identify. 
Nevertheless, these measurements reveal that under normal operating conditions, a minimum of 8.8$\%$ of the applied RF port power is transferred to the tunnel junction. For example, taking the on-resonant case of 312.032 MHz, an applied port power of -27 dBm (10 mV RMS) results in -48 dBm at the junction (0.88 mV RMS). This loss is remarkably smaller compared to what would be expected after transmission through the various RF coax and components shown in Fig.\ \ref{Fig:layout}. However, we note that these relatively low port powers fall into the low temperature regime where the directional coupler does not function according to specifications. In particular, the coupled port which we use for the 'RF-in' path should have an ideal loss of 15 dB. Indeed the reflectometry signal measured along the return path of our tank circuit at low port powers is quite distorted compared to the higher temperature case, see Fig.\ \ref{Fig:pdepLin}(a) for a comparison. Unfortunately, at higher powers where our reflectometry measurements are routinely performed, the coherence peaks broaden beyond any quantifiable measure. As such, this method of transmission measurement is useful in the limit where the applied RF amplitude does not greatly exceed the size of the superconducting gap.

\begin{figure*}[ht]
	\centering
	\includegraphics[width=1\textwidth]{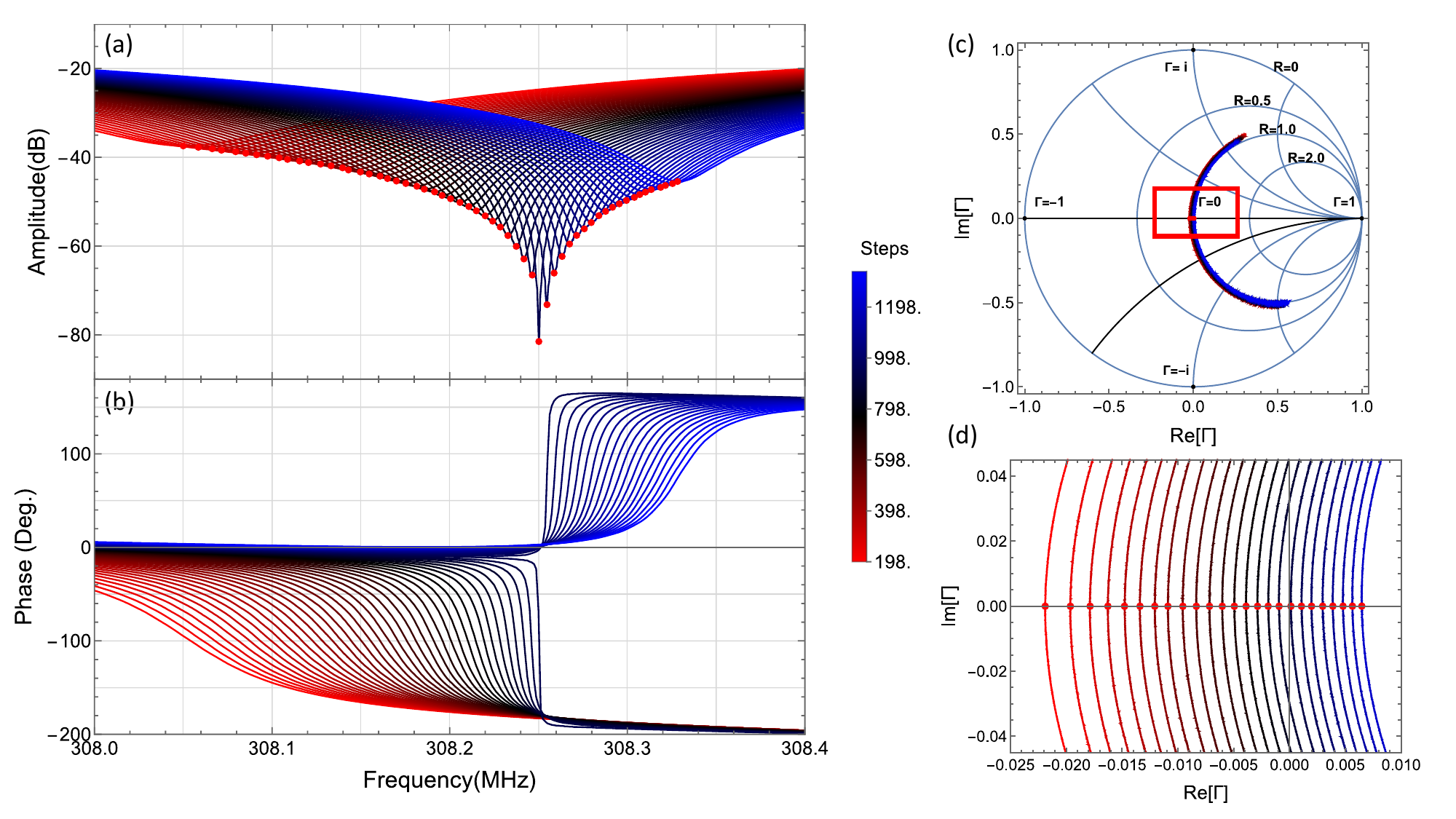}
	\caption{Frequency dependent amplitude (a) and phase (b) reflectometry measurements of the tank circuit as a function of tip-sample distance. The distance is quoted in "steps" of the coarse Z walker, with each step being roughly 30 nm. The red to blue color blends respectively indicate smaller to larger tip-sample distances. Every 10th cure is plotted for clarify. A clear matching point can be identified at 308.25 MHz. (c) Smith chart representation of the reflectometry measurements. Note that an amplitude offset of 2.5 dB has been added to roughly confine the frequency traces to the Re(Z) = 1 arc. Likewise, an overall linear correction to the phase has been added to constrain the resonance positions to the real axis. (d) Expansion of the smith chart in the vicinity of the resonances (red rectangle in c).   }
 \label{Fig:smith}
\end{figure*}
\section*{IV. Resonator sensitivity}
As discussed above in section I, the quality factor of the LC tank circuit provides a relevant metric to quantify the sensitivity of the RF reflectometry. Under the assumption of an ideal RLC circuit where $Q_{int} \propto 1/R$, the resonator was constructed with superconducting components such that the quality factor becomes large below the transition temperature of Nb. However, across multiple cool-downs to base temperature, we observed a wide range of quality factors and final resonance frequencies when the tip-sample distance is within tunneling range. 

To more effectively characterize the resonator,  a standard Au(111) sample was used to perform systematic resonator measurements as a function of the tip-sample distance "z".  Here, z is iteratively varied over a range of 0-1200 steps of the coarse walker, after which a small frequency sweep is taken in the vicinity of the tank circuit resonance at 308 MHz. From a lumped element perspective, changing the tip position in this manner should ideally affect the tip-sample capacitance $C_\mathrm{TS}$ while other parameters (e.g.  $C_\mathrm{Str}$, L, etc...) remain constant; for smaller distances, $C_\mathrm{TS}$  should be relatively large compared to a fully retracted tip. In practice however, it is likely that $C_\mathrm{Str}$ also changes slightly as a function of z. 

In the present set-up, a single coarse step corresponds to a tip-sample distance (\dtip) of roughly 30 nm at base temperature, which we estimate using the z-piezo scanner. The resultant data in Fig.\ \ref{Fig:smith} shows the evolution of the tank circuit resonance at various tip positions, with the starting position (red) close to tunneling distance (i.e. z $<$ 1 nm), while the final retracted tip position (blue) corresponds to z $\sim$ 36 $\mu$m. Indeed, as the tip is retracted from the sample surface, the resonance increases from 308 MHz to 308.33 MHz, suggesting a small decrease in the total capacitance. 

Further observation of the reflected amplitude in Fig.\ \ref{Fig:smith}(a) also reveals that the line shape evolves to a much sharper response after retracting the tip by 846 steps from tunneling distance ($\sim$25$\mu$m) to a resonance frequency of 308.252 MHz. Passing through this matching point, the corresponding phase flips by 180{\textdegree}, after which the apparent quality factor quickly decays back towards a relatively broad line shape as the tip is retracted by a total of 1200 steps. It is important to emphasize that the tank circuit is incredibly sensitive to small temperature fluctuations. As such, in order to accurately obtain a smooth trend of the line shape, the rate at which the STM tip is retracted needs to be kept constant such that heat generated by the piezo motion can reach equilibrium with respect to the cooling power of the mixing chamber. We further note that the second resonance at 423 MHz displayed in Fig.\ \ref{Fig:layout}(c) does not undergo anywhere near as dramatic a change over the full 1200 step retraction and is therefore neglected for the rest of this discussion. 

The apparent trend of the phase and amplitude data clearly demonstrates a progression through the critical coupling condition where, from Eq. 1, $Z_\mathrm{Tot}(\omega_\mathrm{0}) = Z_0$ (i.e. $\Gamma$ = 0). More rigorously, critical coupling can be defined in terms of a simple ratio between the internal and external quality factors ($Q_\mathrm{int}$ and $Q_\mathrm{ext}$) \cite{reviewReflect}:
\begin{equation}
\begin{split}
    \beta=\frac{Q_\mathrm{int}}{Q_\mathrm{ext}} 
\end{split}	\label{eq:beta}
\end{equation} 
where the ideal case for $\beta = $ 1 defines the critical coupling condition. In the context of the approach data, the progression from the $\beta < $ 1 (under-coupled) to the $\beta > $ 1 (over-coupled) regime as a function of tip position is well illustrated by the smith chart representation given in Fig. \ref{Fig:smith}(c). Here, a linear global phase correction is applied to the data such that the resonance positions lie on the real axis of the smith chart. Performing a phase correction in this manner is justified by enforcing that $Im(\Gamma) = 0$ on resonance. The resultant red points given by the resonance positions highlighted in Fig. \ref{Fig:smith}(a), pass through the $\Gamma$ = 0 origin as the tip is retracted from the sample surface. 

To quantify the trends in $Q_\mathrm{int}$ vs $Q_\mathrm{ext}$, the approach data was fit to a generalized reflection model according to ref. \cite{haozhi}:
\begin{equation}
\begin{split}
    S_\mathrm{21}=1- \frac{\frac{Q_\mathrm{r}}{Q_\mathrm{ext}}e^{i\phi}}{1+2iQ_{r}\frac{f-f_0}{f_0}}
\end{split}	\label{eq:S21}
\end{equation} 
Here, $S_\mathrm{21}$ is expressed in terms of the resonator frequency $f_\mathrm{0}$, a constant phase offset $\phi$, and the total loaded quality factor  $Q_\mathrm{r}=(Q_\mathrm{int}^{-1}+Q_\mathrm{ext}^{-1})^{-1}$, with $Q_\mathrm{ext}$ and $Q_\mathrm{int}$ representing the external and internal quality factors. This Lorentzian function was simultaneously fit to the entire series of phase and amplitude response of the approach data displayed in Fig. \ref{Fig:pullback}. See Appendix A for a detailed description of the fitting procedure. The resulting quality factors, displayed in Fig. \ref{FIG:A1}(c) show that $Q_\mathrm{int}$ remains roughly constant at 95, while  $Q_\mathrm{ext}$ linearly increases from 92 to 96 as the STM is withdrawn from the sample surface. These variations in $Q$ correspond to a change from $\beta \approx $ 1.03 to  $\beta \approx $ 0.98. Although this variation in $\beta$ is quite small throughout the entire range of the tip retraction, there is a significant influence on the overall line shape. 
These results demonstrate that the current tank circuit configuration yields an unoptimized coupling constant at tunneling distance. From this, it is evident that achieving proper impedance matching at tunneling distance represents a significant experimental challenge that needs to be overcome in order for this technique to be successful. Unfortunately, the fitted results from Eq. \ref{eq:S21} do not map well to any particularly obvious lumped element model. We attribute the difficulties in modeling to a poor understanding of the parasitic sources of capacitance and inductance, particularly in the vicinity of the tip plate. For example, we have observed that subtle changes to the orientation of the bias plate can shift the resonance frequency by 2-3 MHz, which can drastically alter the matching conditions at base temperature. Despite our incomplete understanding of the exact contributions to the final loaded Q, the trend in line shape suggests that decreasing the capacitive coupling at tunneling distance will push the resonance towards the optimal matching point. 

In the context of quantum dot devices employing dispersive readout, there are multiple examples of tunable matching networks that employ varactors capable of operating at cryogenic temperatures\cite{varactor,natureVaractor}. In microwave impedance microscopy set ups, impedance matching is instead achieved using a $\lambda/4 $ stub tuner as briefly outlined in ref. \cite{MIM_REV}. Future design iterations in our system will likely need to incorporate similar designs to reliably tune to maximum sensitivity. However, in practice, we have observed that opting for longer STM tip lengths tends to enhance Q since it maximizes the distance from the components on the tip plate to the sample. By maximizing this distance, we can reliably reduce parallel sources of parasitic capacitance. However, this does not represent a reliable/universal tuning method to enhance Q.

\begin{figure*}[ht]
	\centering
	\includegraphics[width=1\textwidth]{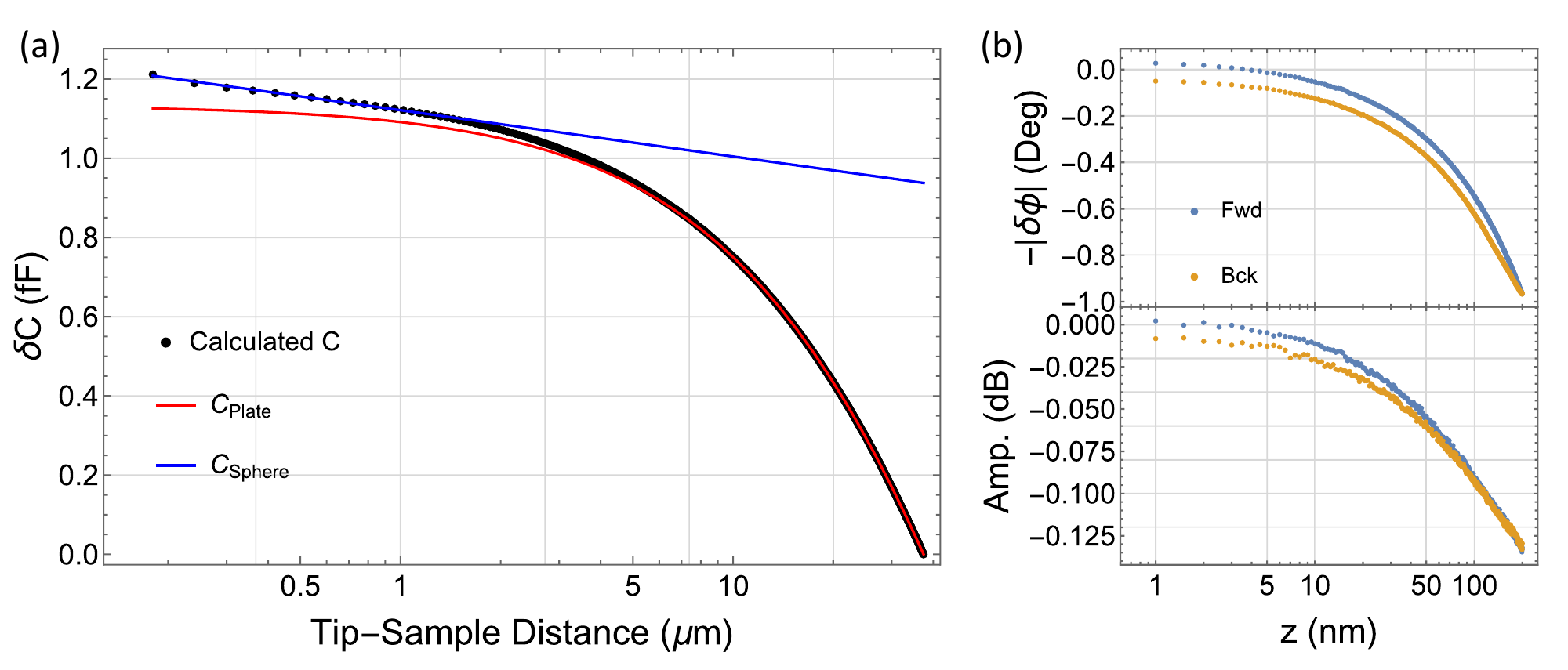}
	\caption{(a) Calculated Capacitance assuming a series RLC with inductance of 470 nH (see main text for discussion on assumptions). The capacitance is plotted as an offset with respect to the tip fully retracted (approximately .57 pF). Note that the data are evenly sample with respect to z (i.e. not ln(z)). (b) Phase and (c) amplitude shifts measured with respect to tip-sample distance using the z-piezo.}
 \label{Fig:pullback}
\end{figure*}

Though the exact value of capacitance as a function of the tip-sample separation is unable to be determined from the pull back data, we can approximate values of $C_\mathrm{TS}$ by assuming a series RLC with the calculated value of $L$ = 470 nH based on the coil dimensions described in section II. The resulting calculated values are shown by the black data points in Fig.\ \ref{Fig:pullback} as a function of ln(z). Since we are primarily interested in variations with distance, the capacitance is plotted as the relative change $\delta$C with respect to the capacitance at z = 36 $\mu$m, when the tip is full retracted. In this approximation, the change in resonance frequency from 308 MHz to 308.33 MHz over the full coarse walker range corresponds to a series capacitance variation of 1.2 fF. 

 Two notable trends in  $\delta$C vs ln(z) can be observed: for z $<$ 1.5 $\mu$m, the capacitance follows a clear logarithmic trend (i.e. linear when plotted against ln(z) ), while for larger distances, $\delta$C rapidly declines. At these larger distances, a simple parallel plate capacitor model replicates the data quite well, given by the red curve in  Fig.\ \ref{Fig:pullback}. Here, an arbitrarily far cut off point at z $>$ 10 $\mu$m was fit to a simple $A\epsilon_0/z$ law, resulting in a cross sectional area of 42.1(1) $\mu$m$^2$. For a circular cross section, this area corresponds to a diameter of 7.32 $\mu$m, which is considerably smaller than the diameter of the Nb wire from which the tip is etched (250 $\mu$m). The fitted red curve replicates the data up to a tip-sample distance of 3 $\mu$m where an obvious divergence can be observed. To account for the logarithmic behavior for smaller z, a spherical capacitor approximation was used for which $C(z) = 2\pi\epsilon_0R_\mathrm{eff}$ln$(8R_\mathrm{eff}/z)$ is valid in the limit where $z < R$ \cite{bestTipCap}. However, due to the fact that the data is not well sampled below $z < $1 $\mu$m, we instead choose a data cut off that minimizes fit error while still ensuring $R_\mathrm{eff} < z$. This approach results in a fit value of $R_\mathrm{eff}$ = 866(26) nm using only the first 12 data points. 

The $\delta$C vs z behavior we observe is consistent with similar studies where capacitance is measured for adjustable-sized vacuum gaps in both STM and scanning capacitance microscope geometries \cite{conePaper}. For the best point of comparison in a STM context, in ref. \cite{bestTipCap} $\delta$C vs z was directly measured between the STM tip and sample using a capacitance bridge for a series of sharp and blunt tips. Focusing on the 'close' regime, a tip of radius 2000 nm as measured via SEM resulted in a capacitance variation of roughly 0.3 fF over a distance of 1 $\mu$m, whereas we calculate an approximate capacitance change of 0.25 fF. Likewise, a similar fit to the logarithmic approximation yielded $R_\mathrm{eff}$ = 2400 nm (i.e. somewhat close to what was inferred via SEM). We further note that this simple spherical fit provides the best and most realistic fit errors compared to the lengthier conical expressions that can be found in refs. \cite{capacitiveForce, conePaper}.From these observations, we can conclude that although $C_\mathrm{TS}$ can not be directly measured using our integrated tank circuit, the approximate calculated values provide a reasonable estimate for the physical geometry of the STM tip. 

An alternative method to probe tip-sample capacitance can be achieved by repeating similar pull-back measurements as in Fig.\ \ref{Fig:smith} using the fine z-piezo scanner. The resultant phase and amplitude change at constant frequency over a close range of 200 nm from the sample surface are respectively shown in Fig.\ \ref{Fig:pullback} (b) and (c) against a log scale of z. Even for very short distances, the phase appears to follow an extremely similar trend to the pull back data in Fig.\ \ref{Fig:smith}. With this in mind, an attempt was made to re-fit the data following a similar derivation for Eq. \ref{eq:dgdc} to describe phase variations as a function of capacitance $\partial \phi/\partial C =2Z_0 R/(R^2-Z_0^2) (Q/C)$. Changing variables from $\partial \phi/\partial C$ to $\partial \phi/\partial z$ according to the $C(z)$ dependence described above for a spherical capacitor yields the relation that $\phi(z) \propto\\ $ln$($ln$(8R_\mathrm{eff}/z)$ assuming the frequency is tuned to resonance. However, fits to the data in \ref{Fig:pullback}(b) were not successful as the trend in $\phi$ does not follow a clear double log behavior. This unsuccessful attempt at modeling is likely due to the following: i. our tank circuit is not an ideal LCR, ii. the resonance drifts slightly as the tip is retracted, and iii. more than one effective lumped element parameter depends on the tip-sample distance. While the conclusions in i. and ii. are well supported by previous discussions, the relatively large change in amplitude as a function of distance in Fig.\ \ref{Fig:pullback} gives strong support for iii. We specifically investigate amplitude dependence in the proceeding section as a way to probe local sample surface variations.

\section*{V. Imaging using RF Reflectometry}
    During a particular set of measurements on a Au(111) standard, it was found that ice had condensed on the sample surface. This water condensation unintentionally occurred when transferring the sample from the preparation chamber down into the dilution refrigerator. Although the resulting sample surface did not yield the best environment for tip preparation, it serendipitously provided an excellent test surface for our integrated RF capabilities. In particular, we note that while the dielectric permittivity of ice has not been explicitly measured at mK temperatures within a bandwidth of 300 MHz, we can expect $\epsilon \approx$ 3 based on previously measured temperature and frequency dependent studies of water ice \cite{icedielectric}. 

    \begin{figure*}[ht]
	\centering
	\includegraphics[width=1\textwidth]{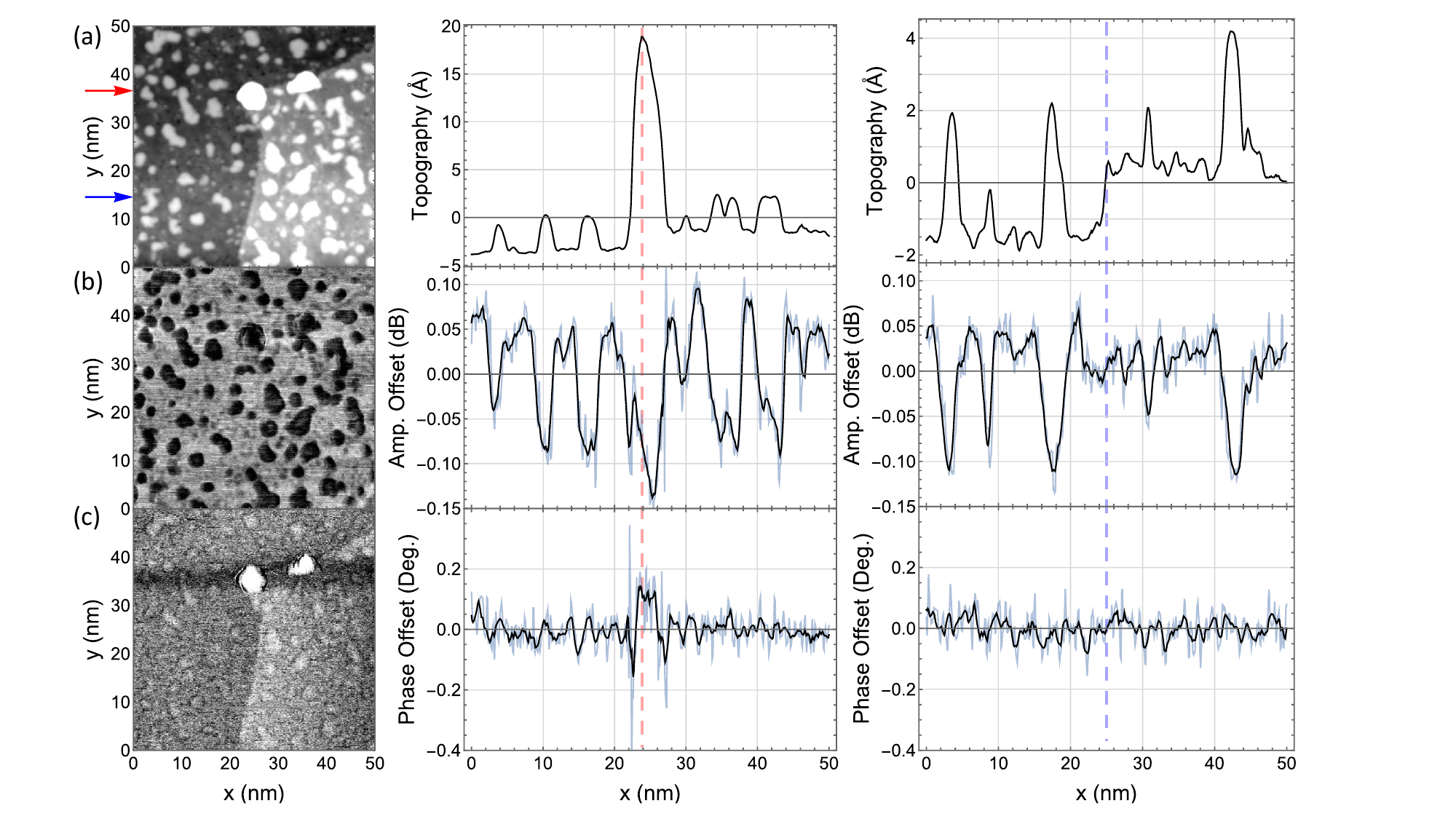}
	\caption{(a) 50x50 nm (512x512 resolution) topography image of Au(111) containing two separate terraces of the island features discussed in the main text. All topography data is subject to a 2D plane fit. The middle and right-most panels respectively correspond to the cross sections taken along the trajectory of the red (y = 36.6 nm) and blue (y = 14.6 nm) arrows). The red and blue vertical dashed lines respectively highlight the topography associated with the high particle (middle) and step edge (right). (b) Amplitude response of the tank circuit, which is measured with the topography simultaneously. The amplitude data is subject to a line fit correction to account for slow drift before being decorrelated with respect to the tunneling current. (c) Phase response of the tank circuit after decorrelating with respect to tunneling current and topography. Refer to the appendices for a description of the decorrelation procedures. For the cross sections of the RF data in the middle and right panels, raw data is shown in light blue, while the bolder black traces are given by a 5-pt moving average.}
\label{Fig:topo}
\end{figure*}

\begin{figure*}[ht]
	\centering
	\includegraphics[width=1\textwidth]{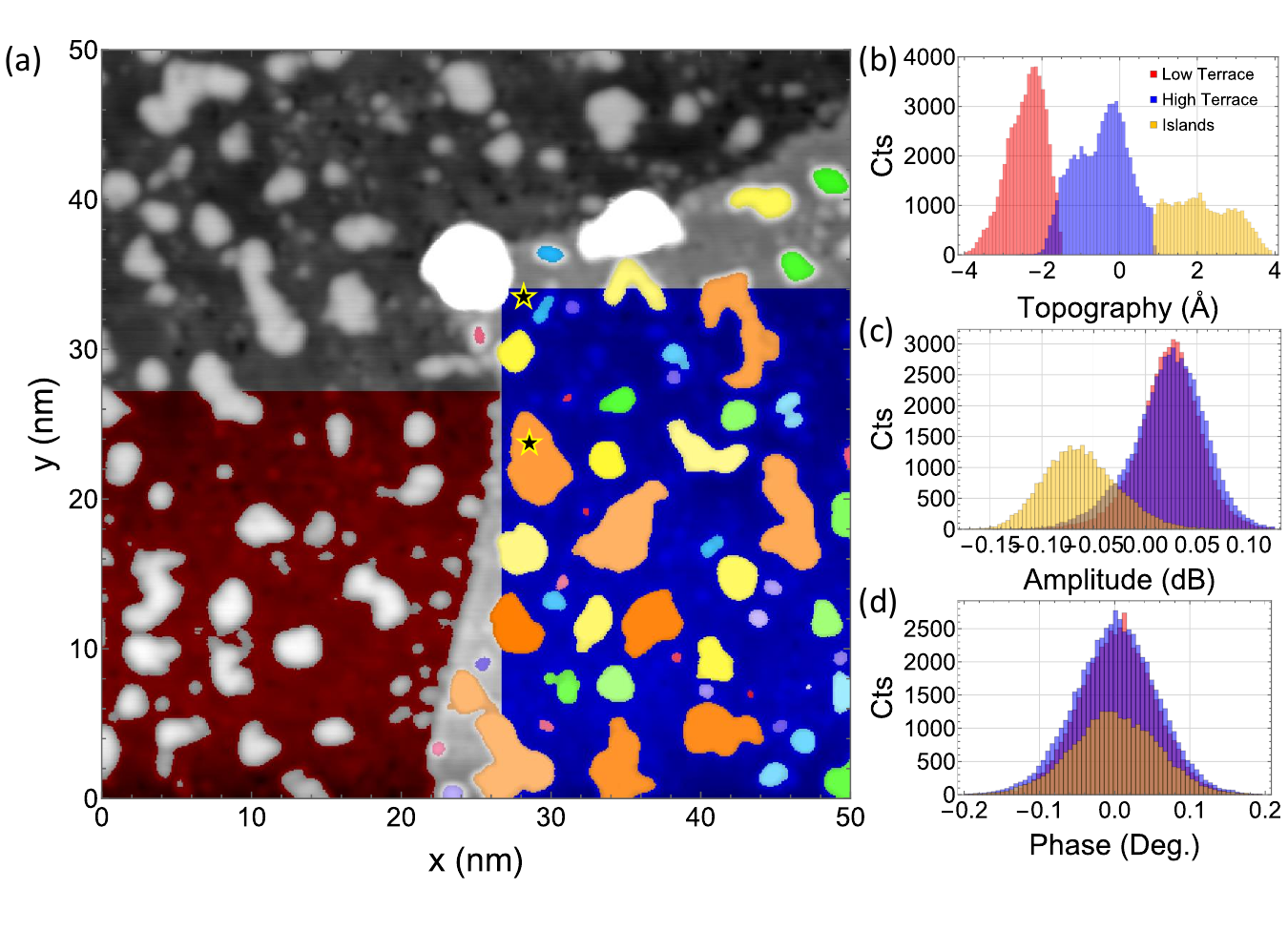}
	\caption{(a) Morphological components of the topography image showing different data selections. The red plane selects for the lower Au(111) terrace which neglects island-like features. The blue plane selects for the upper Au(111) terrace. Both plane components have been trimmed such that there is no spatial overlap, and to avoid the upper (darker) part of the scan image where the plane fit correction is less effective. The islands in the upper terrace are highlighted by multi-colored selections. The two largest particles have been neglected. The distributions corresponding to the black stars are further scrutinized below in Fig. \ref{Fig:Distributions}. The topography, amplitude and phase distributions associated with these selections respectively shown in (b), (c) and (d)}

 \label{Fig:islands}
\end{figure*}

    The resulting topography at base temperature after subjecting the Au(111) sample to the conditions described above is shown in the 50 x 50 nm$^2$ image in the left panel of Fig.\ \ref{Fig:topo}(a). Here, a number of scattered, irregularly sized islands can be observed on the Au(111) surface with a partial surface coverage. We attribute the vast majority of these structures to the presence of amorphous ice, with the exception of the two largest particles at (x,y) = (22 nm, 37 nm) and (35 nm, 38 nm). These two particles are distinctly higher than the rest of the surface features, and notably pin the step edge between two terraces present in the topography. The smaller scattered islands have maximum heights of roughly 2-4 \AA\ above a plane fit corrected baseline. For a point of comparison, these height variations are in line with previous AFM/STM studies in which ordered hexagonal ice islands of 2.5 \AA\ were carefully grown on a Au(111) surface at 120 K \cite{iceOnAu}. However, the absence of well-defined islands with sharp crystal facets in the data presented here likely suggests that they formed atop a so-called 'wetting layer' rather than a bare Au(111) surface, though this is difficult to confirm from topography alone.  
    
    The abundance of different topographical features provides a convenient test bed to characterize the response of our resonator. Fig.\ \ref{Fig:topo}(b) and (c) respectively show the corresponding amplitude and phase of the reflectometry as a function of tip position with the tank circuit fixed to a resonance of $f_0=$ 308~MHz (i.e. the leftmost red curve in Fig.\ \ref{Fig:smith}). In order to minimize the effects from cross talk between tunneling current and the RF measurements, all data is subjected to a decorrelation procedure (see appendix for more details). We note here however that the phase response is decorrelated with respect to the tunnel current \textit{and} topography, while the amplitude was only decorrelated against the tunneling current.
    
    Qualitative analysis of the post-processed RF images reveals a strong anti-correlation between amplitude and topography (i.e. the islands appear dark in amplitude) while a weak positive correlation between topography and phase remain (island structures tend to appear bright in phase). To explore these trends in more detail, two different cross sections taken along the trajectory of the red and blue arrows of Fig.\ \ref{Fig:topo}(a) are respectively shown in the middle and right panels. These cross sections were selected to highlight comparisons of the RF response of the ice islands with respect to: i. the large particle at the step edge (middle panel) and ii. the single atomic step which define the two gold terraces (right panel). First focusing on the amplitude cross sections, the peaks in the topography manifest as apparent dips on the order of -0.1 dB. These amplitude variations are straightforward to explain: the presence of ice on the sample surface increases the relative permittivity, which gives rise to an increase in absorption. However, what is most interesting is that the relative amplitude variations do not necessarily scale proportionally with the topography. For instance, the cross section in the middle panel shows consistent amplitude dips all of a similar 0.1 dB magnitude despite the fact that the high particle located at x = 26 nm (vertical red dashed line) is nearly 4 times taller than the smaller island profiles. This difference in absorption indicates that the large particles in Fig.\ \ref{Fig:topo}(a) are distinct from the rest of the smaller/scattered islands; they are more reflective than the ice islands, but less reflective than the bare Au(111) surface. In contrast, variations among the smaller ice islands correlate well with their topography—e.g., taller ice particles exhibit proportionally deeper dips in relative amplitude. Somewhat counterintuitive to this, while differences in absorption between ice islands are clearly resolved, the reflected amplitude appears completely insensitive to topograhpic variations on the actual gold surface. This insensitivity is clearly illustrated by the amplitude cross section in the right most panel which is taken along the trajectory of the blue arrow in Fig.\ \ref{Fig:topo}(a). The amplitude profile in the vicinity of the single atomic step (2.4 \AA\ ) at x = 25 nm (blue dashed line) reveals no discernible shift in baseline. Indeed, careful observation of Fig. \ref{Fig:topo} (b) reveals almost no indication of two separate terraces (or any other surface related geometry) in the scan area. 
   
In contrast to the strong amplitude response as a function of the tip position, the phase response is markedly weaker. From the two cross sections in Fig\ \ref{Fig:topo}(c), the only detectable change in phase occurs in the vicinity of the large particle. It is unclear whether this weak phase shift results from an actual change in tip-sample capacitance, or if it is a remnant of an imperfect decorrelation procedure (see appendix). Nevertheless, the weak contrast is unsurprising given the previous discussion of the pull back data in section IV. Specifically, for a change in permittivity of $\epsilon = 1 \rightarrow 3$ (i.e. from gold to ice), we expect a total phase shift of $\sim$0.1 mDeg based on the calculated scaling factor for an ideal RLC circuit. Such a small phase change would require at least an order of magnitude improvement in SNR to be detectable. We further note that since $\Delta\phi$ is linearly proportional to $Q$, this SNR is certainly achievable provided the impedance can be properly matched at tunneling distance. As such, this discussion illustrates the importance of achieving the best Q possible in the context of successfully resolving tip-sample capacitance. 

To support the sensitivity trends in amplitude and phase, we compile the distributions associated with different topographical features by grouping sets of morphological components in Fig.\ \ref{Fig:topo} (a). To constrain the associated components, the topography image was binarized at different thresholds to selectively group different sets of features, after which the thresholded selections were piped to Mathematica's standard morphological component functions \cite{mathematicawebsite}. Fig.\ \ref{Fig:islands}(a) shows the three relevant selections of components that include: i.~the lower terrace neglecting island-like features (red), ii. the upper terrace, also neglecting any island selections (blue), and iii. an expanded complement of (ii.) which includes all relevant surface features on the upper terrace with the exception of the two high particles. The distribution of the corresponding selections for the topography, amplitude, and phase variations are respectively shown in Fig.\ \ref{Fig:islands}(b-d). Note that an overall offset has been applied to the topography such that the mean of the blue selection corresponds to a height of 0 \AA; no analogous offsets have been introduced into the RF reflectometry data here. From the results plotted in Fig.\ \ref{Fig:islands}(c), the red and blue distributions corresponding to the low and high terrace selections exhibit no detectable shift in the amplitude despite there being a relative height difference of 2.4 \AA. However, for the selections of the island-like features (light orange), there is a very obvious amplitude shift of the distributions, where $\Delta \mu_\mathrm{amp}$  = 85 mdB according to Gaussian fits.  This observation is consistent with the cross section displayed in the right panel of Fig.\ \ref{Fig:topo}(b). Evidently the amplitude response of the tank circuit is far more sensitive to changes in total absorption rather than topographical variations within one material. Compared to amplitude, the de-correlated phase distribution associated with the island selections in Fig.\ \ref{Fig:islands}(d) shows an extremely small shift in mean compared to the selection of the terrace on the order of 1 mDeg. However, this shift is still within the FWHM of 55 mDeg. 

Further analysis of the phase and amplitude response in the vicinity of each colored island is described in Appendix C. In particular, compiled distribution charts of the amplitude response in Fig.\ \ref{Fig:Distributions} (b) show that amplitude sensitivity is limited down to a minimum island size of 5 nm$^2$. 
    
\section*{Conclusions and future work}
We have described the successful implementation of an RF-STM that is capable of sensitive reflectometry measurements at a resonant frequency of 300 MHz. The electron temperature and transmission have been characterized by careful measurement of the coherence peaks afforded by a Nb tip. Tip-sample distance variations reveal that proper impedance matching ultimately dictates the sensitivity of this technique. Even with a diminished quality factor at tunneling distance, we can successfully resolve amplitude variations of island-like features of crystalline ice on Au(111) for island sizes above 5 nm$^2$. Phase variations remain weak due to our relative insensitivity to tip-sample capacitance for constant z. 

Future efforts will focus on enhancing sensitivity through the integration of reliable and repeatable tunable matching networks, which will improve the quality factor of the tank circuit. Ultimately, our goal is to leverage this technique to investigate novel semiconductor materials that have relevance to the QIS community.

\section*{Acknowledgements}
We thank Haozhi Wang and Neda Foroozani for their discussions and advice on fitting the resonator responses in section IV.

\appendix
\renewcommand\thefigure{\thesection.\arabic{figure}}
\section{Fitting Approach Data}
\setcounter{figure}{0}

\begin{figure*}[ht]
	\centering
        \includegraphics[width=1\textwidth]{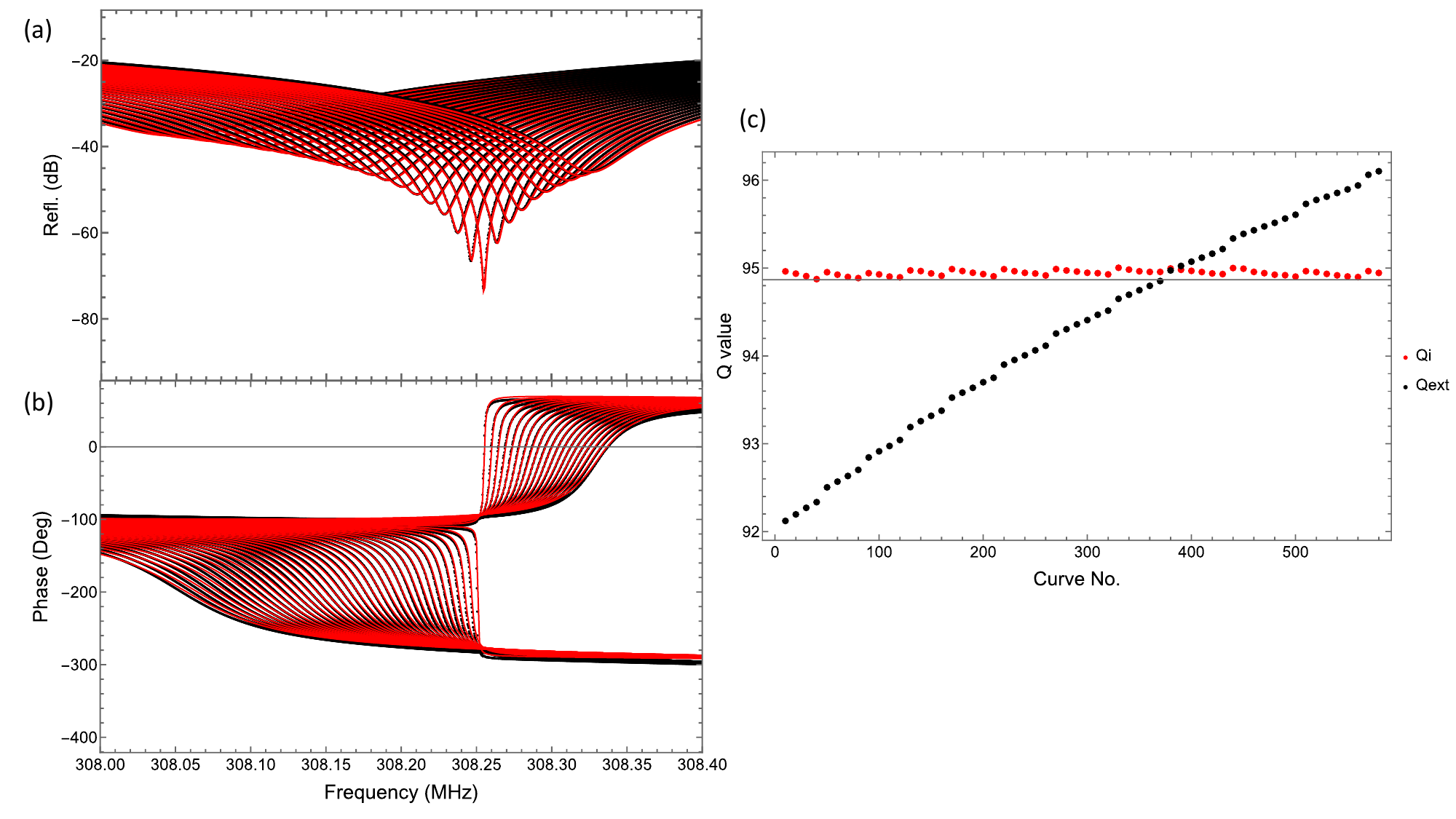}
	\caption{Fitted amplitude (a) and phase (b) in red plotted with respect to the data in black. (c) Fit results $Q_{ext}$ (Black) and $Q_{int}$ (Red). 
 \label{FIG:A1}
 }
\end{figure*}

The full procedure used to fit the approach data shown in main text Figure \ref{Fig:smith} is described in the appendix of the thesis in reference \cite{gaoThesis}. In particular, the function given by the red fits in Figure\ \ref{FIG:A1} includes an additional complex-valued scaling prefactor and a constant phase offset to Eq. \ref{eq:S21}. Before fitting the reflectometry, the data was transformed to the real and complex 'I/Q' quadrature components. The curves were then resampled such that the data points in the vicinity of the resonance frequency were given a higher weight. The quadrature components of all 58 data sets (i.e., every 10th curve) were simultaneously fit such that the phase and scaling factors serve as global parameters, while the quality factors $Q_{ext}$ and $Q_{int}$ were allowed to vary for each frequency response. After transforming back to magnitude (dB) and phase (deg), the corresponding fits are plotted with respect to the approach data in Fig.\ \ref{FIG:A1} (a) and (b). Notably, the fitted quality factors in Fig.\ \ref{FIG:A1} (c) replicate the critical coupling condition where $Q_{int}$ = $Q_{ext}$. We note that the small 4-point modulation in the final fit results arises from the method in which the data was re-sampled. 

\section{Decorrelation Procedure}
\setcounter{figure}{0}

\begin{figure*}[ht]
	\centering
    \includegraphics[width=1\textwidth]{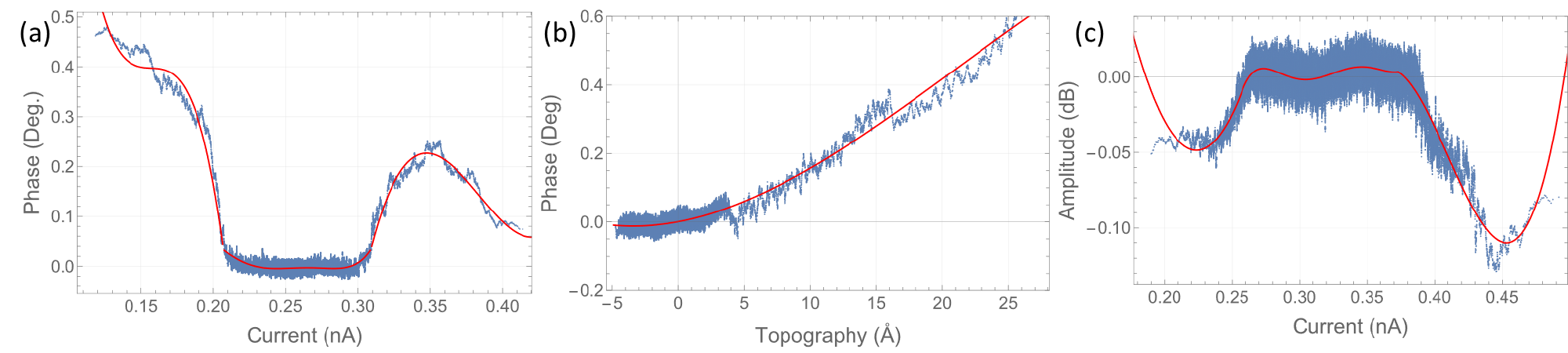}
	\caption{(a) Smoothed phase vs tunneling current data used to perform the decorrelation described in the appendix. The red trace corresponds to a piece wise function which consists of three separate fifth order polynomials. The boundaries for the polynomials are determined by 'dense' data region between 200 and 300 pA. (b) Smoothed current decorrelated phase data vs Topography. The red trace is a simple 3rd order polynomial. (c) Smoothed amplitude vs tunneling current data along with a similar piece function as employed in (a). 
 \label{FIG: A2}
 }
\end{figure*}

Here we briefly describe the procedure used to carry out the decorrelation of phase and amplitude data discussed in main text Figures\ \ref{Fig:topo}, \ref{Fig:islands}, \ref{Fig:Distributions}. We perform this decorrelation to ensure that cross-talk between the AC and DC channels of the cryogenic bias tee is minimized. Likewise, we also wish to minimize correlations between amplitude variations and tunnel resistance, as has been reported in previous RF-STM measurements \cite{nat2007}. First addressing the phase response, the 512x512 matrix is unwrapped into a single 1D array and plotted against the corresponding tunneling current. To compensate for noise, the data is sorted and smoothed using a 100-pt moving average. The resulting smoothed data are fit to a fifth order piece wise function, with the fit boundaries being given by the cut-offs of the 'dense' data between 200 and 300 pA as shown in Fig. \ref{FIG: A2}(a). After subtracting this fitted correlation, the procedure is repeated with respect to the topography using a simple 3rd order polynomial function as  shown in Fig.\  \ref{FIG: A2}(b). Note that the sparser data above 3.6 \AA\ corresponds to the high particles which pin the step edges of the terrace shown most clearly in Fig.\ \ref{Fig:topo}(a). Concerning the amplitude data, a similar piece wise function as described for the phase vs tunneling current is employed. The smoothed data along with the fit in red is shown in Fig.  \ref{FIG: A2}(c). For the amplitude response, only correlating with respect to the tunneling current proved to be effective; introducing a topography decorrelation into the amplitude response resulted in a non-physical distortion of the distributions shown in Fig.\ \ref{Fig:Distributions}(b) and was therefore neglected in all analysis. 

\section{Limits on RF sensitivity with respect to island size}
\setcounter{figure}{0}
\begin{figure*}[ht]
	\centering
	\includegraphics[width=1\textwidth]{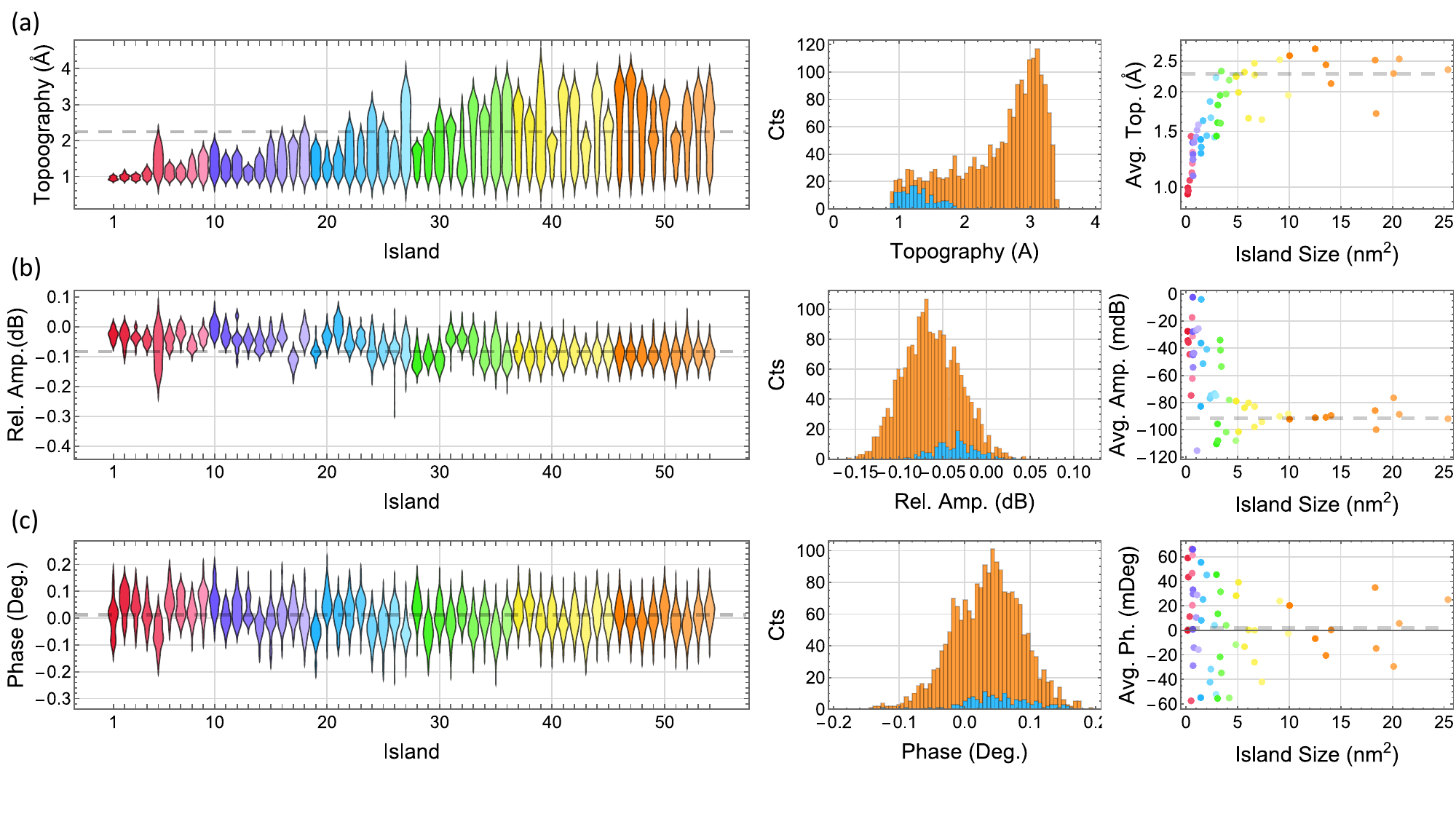}
	\caption{Distribution charts for the topography (a), relative 
 amplitude (b) and phase (c) according to the multi-colored island selections in Fig.\ \ref{Fig:islands}(a). All islands are sorted by size (i.e. surface area), with island 1 being the smallest selection (red) to island 54 being the largest (orange). The distribution charts themselves are a modified version of a so-called 'box and whiskers' plot, where the peaks within a given distribution is represented by the relative width. More explicitly, distributions with multiple 'bulges' correspond to distributions with multiple peaks \cite{mathematicaDist}. The middle panel shows two distributions in detail associated with islands labeled 20 (blue) and 50 (orange), which are given by the positions of the starred features in \ref{Fig:topo}(a). The far right panel shows the average value associated with each distribution as a function of island size. The horizontal dashed lines are given by the overall average for island sizes greater than 5 nm$^2$.
 \label{Fig:Distributions}
 }
\end{figure*}

    To gain further intuition into the reflectometry response with respect to variations in size, we break the island selections down into their individual components and calculate their corresponding distributions in Fig.\ \ref{Fig:Distributions}(a-c). Here, all distribution charts are sorted by the surface area of a given component selection, which we parameterize as 'island-size'. Note that all components with surface area smaller than 0.1 nm$^2$ have been neglected from the analysis. From the topography distributions, we see that as the island size increases, so too does the average height (left panel, log scale) up to an island size of 5 nm$^2$ (yellow and orange island selections in Fig.\  \ref{Fig:islands}a). Above this soft threshold, the average topography fluctuates around 2.2(6) \AA. This trend in height suggests that the islands tend to grow bigger instead of taller within a single mono-layer. The dependence of topography on island size is well reflected by the relative amplitude shifts; for island size $>$ 5 nm$^2$ , the mean amplitude shift converges to -89(8) mdB. However, below this threshold the amplitude distributions appear to somewhat randomly approach a relative offset of 0 dB. This data suggests the amplitude response is only sensitive up to a certain island size. To test this claim, we closely inspect two islands on either side of the 5 nm$^2$ size threshold and plot their respective distributions in the middle of Fig.~\ref{Fig:Distributions}: island $\#$50 (20 nm$^2$, orange) and island $\#$20 (1.5 nm$^2$, blue). The positions of these islands are marked in \mbox{Fig.~\ref{Fig:islands}(a)} by the black stars. Focusing on the orange distribution, we see that the topography distribution is significantly more asymmetric (i.e. it is peaked at 3 \AA) than the amplitude distribution. Given the results summarized in Fig.~\ref{Fig:Distributions}(c), this technique is inherently insensitive to topographical variations within one type of material. This suggests that the absorption due to the presence of the ice islands is significantly more sensitive to surface area than to height. Though the threshold for minimum island size is on the order of 5 nm$^2$, we note that this is extremely small compared to the estimated tip radius as discussed in the previous section. This suggests that a sharp tip is not necessary to obtain sufficient spatial resolution for RF loss imaging. 
    
    For completeness, Fig.~\ref{Fig:Distributions}(c) shows the phase distributions associated with the same set of island selections in (a) and (b). Given the observations presented above, it is unsurprising that there are no discernible trends among the distributions apart from the fact that the overall means plotted in the far right  seems to converge towards $\phi$ = 0 for larger island size. However, it is unclear whether this convergence is due to the remnants of uncompensated signal after the decorrelation procedure outlined in the appendix below. As such, we cannot conclude anything meaningful from the phase data. 

\clearpage


\bibliographystyle{apsrev4-2}
\bibliography{sample}

\end{document}